\documentclass{mpe_report}

\usepackage{psfig,graphicx,epsfig}
\usepackage{color}
\usepackage{amsmath,amssymb,epic,eepic,array}

\begin{document}

\pagenumbering{arabic}
\setcounter{page}{137}

 \renewcommand{\FirstPageOfPaper }{137}\renewcommand{\LastPageOfPaper }{140}
\title{Computation of Neutron Star Surface Emission Spectra 
for Arbitrary Magnetic Field Directions without Diffusion Approximation}
\author{Lun-Wen Yeh\inst{1}, Gwan-Ting Chen\inst{1} 
\and Hsiang-Kuang Chang\inst{1,2}}  
\institute{Department of Physics, National Tsing Hua University, Hsinchu 30013, 
Taiwan, Republic of China
\and Institute of Astronomy,
National Tsing Hua University, Hsinchu 30013, 
Taiwan, Republic of China}
\titlerunning{Computation of Neutron Star Surface Emission Spectra}
\authorrunning{L.-W. Yeh, G.-T. Chen and H.-K. Chang}
\maketitle

\begin{abstract}
To derive physical properties 
of the neutron star surface with observed spectra, 
a realistic model spectrum of neutron star surface emission is essential. 
Limited by computing resources, a full computation of the radiative transfer equations
without the diffusion approximation has been conducted up to date only
for the case of local magnetic
fields being perpendicular to the stellar surface. 
In this paper we report the full-computation result
for an arbitrary field direction.
For comparison we also compute the radiative transfer equation 
using the diffusion approximation. 
For a given effective temperature, the computed spectrum with the diffusion
approximation is always softer than that of a full computation at a non-negligible level. 
It leads to an
over-estimate of the effective temperature if the diffusion approximation 
spectrum is employed in the spectral fitting.
Other characteristics for different magnetic field orientations, such as
the beaming pattern of the two polarization modes and the structure of the atmosphere,
are also discussed. 
\end{abstract}

\section{Introduction}

The spectra of some radio pulsars and radio-quiet isolated neutrons reveal a 
thermal origin in the soft X-ray band (e.g. Becker \& Tr\"umper 1997; 
Pavlov, Zavlin \& Sanwal 2002). 
These thermal emissions are believed to originate from the surface of neutron stars 
and they carry the information of physical properties of the neutron stars. 
Modeling the atmosphere of neutron stars to obtain a more realistic spectrum 
of the surface emission to compare with observations 
may help us to have better knowledge about those properties 
such as the composition, the mass-to-radius ratio and 
the strength of the magnetic field at the surface. 
Many neutron star atmosphere models have been developed in various physical conditions 
(see e.g. Zavlin \& Pavlov 2002, for a review). 
We construct an atmosphere model with pure hydrogen gas. 
In the field range of $10^{11}$-$10^{13}$ G and the effective temperature of 
several million Kelvin, the hydrogen gas can be well approximated as a fully ionized plasma 
(Lai \& Salpeter 1997). 
Therefore only the scattering and the free-free processes are involved in our computation. 
We solve the radiative transfer equation of angle-dependent specific intensity 
for the two polarization modes in a plasma. The magnetic field direction is allowed to be
arbitrary. 
Up to date, such a full computation without the diffusion approximation
has been done only for the case of the magnetic field being
perpendicular to the stellar surface. In the following sections we briefly describe
the method we use and then discuss the results for different field orientation and the
difference between the full computation and that with the diffusion approximation.

\section{Assumptions and the radiative transfer equations}
  
The equation of state adopted for the fully ionized hydrogen plasma is 
that of an ideal gas,
$P\simeq\frac{2\rho k T}{m_{\rm p}}$, where $P$, $\rho$, and $T$ are 
the pressure, density and temperature of the hydrogen gas respectively; 
$m_{\rm p}$ is the proton mass. 
Because the scale height of the neutron star atmosphere ($\simeq 1{\rm cm}$) 
is much smaller than 
the radius of the neutron star ($\simeq 10^6$cm), 
the atmosphere is treated as a plane-parallel surface. 
We assume the atmosphere is in hydrostatic equilibrium,
and all physical quantities are independent of time.
The hydrostatic equilibrium in the atmosphere
is described by the Oppenheimer-Volkoff equation,
\begin{equation}
\frac{{\rm d}P}{{\rm d}z}=
\frac{-\rho GM}{R^2}(1+\frac{P}{\rho c^2})
(1+\frac{4\pi R^3P}{Mc^2})(1-\frac{2GM}{Rc^2})^{-1}
\,\,\, ,
\end{equation}
where $z$ is the height in the atmosphere, $M$ is the mass of the neutron star, 
$G$ is the gravitational constant, and $R$ is the stellar radius. 
The terms in the first two parentheses on the right hand side of equation (1) are very close to unity. 
Equation (1) can be written as
\begin{equation}
\frac{{\rm d}P}{{\rm d}\tau}=\frac{g^*}{\kappa_{\rm sc}}
\,\,\, ,
\end{equation}
where $g^*\equiv\frac{GM}{R^2}(1-\frac{2GM}{Rc^2})^{-1}$,
$\tau$ is the Thomson optical depth defined by ${\rm d}\tau\equiv-\rho\kappa_{\rm sc}{\rm d}z$,
and $\kappa_{\rm sc}$ is the zero-field Thomson scattering opacity.

The transport of photons is described by the radiative transfer equation, which is
\begin{eqnarray}
\mu\frac{{\rm d}I_{\nu}^{i}}{{\rm d}\tau}=\frac{\kappa_{\rm sc}^{i}+\kappa_{\rm ff}^{*i}
}{\kappa_{\rm sc}}I_{\nu}^{i}-\frac{\kappa_{\rm ff}^{*i}
}{\kappa_{\rm sc}}\frac{B_{\nu}}{2} \nonumber\\
 -\frac{1}{\kappa_{\rm sc}}\sum_{{j}'=1}^{2}\oint \frac{{\rm d}\kappa_{\rm sc}^{ij'}}{{\rm d}\Omega}
(\hat{k}\leftarrow\hat{k'})I_{\nu}^{j'} d\Omega^\prime
\,\,\, ,
\end{eqnarray}
where the specific intensity $I_{\nu}^{i}(\theta_{\rm R},\phi_{\rm R})$ is a function of direction
specified by the angle $\theta_{\rm R}$ between the surface normal and 
the propagation of the photon and the azimuthal angle $\phi_{\rm R}$ 
($\phi_{\rm R}=0^\circ$ in the direction of the magnetic field). The two polarization eigen modes are
denoted with the superscript $i$ 
with $i=1$ for the extraordinary (X) mode and $i=2$ for the ordinary (O) mode.
The other notations are that
$\mu\equiv\cos\theta_R$, $\tau$ is the Thomson optical depth,
$B_{\nu}$ is the Planck function, $\kappa_{\rm sc}^i$ is the Thomson scattering opacity 
for the $i$-mode photons, and
$\kappa_{\rm ff}^{*i}=\kappa_{\rm ff}^i(1-\exp{(\frac{-\hbar\omega}{kT})})$ 
is the reduced opacity of the thermal bremsstrahlung for the $i$-mode photons. 
All the opacities can be found, for example, in M\'esz\'aros (1992).
The two boundary conditions for the above equation are
\begin{eqnarray}
I_{\nu}^{i}(\tau=0,\mu,\phi_{\rm R})=0\;\; \mbox{for}\;\; -1\leq\mu\leq0, \nonumber\\
I_{\nu}^{i}(\tau=\infty,\mu,\phi_{\rm R})=\frac{1}{2}(B_{\nu}+\mu\frac{{\rm d}B_{\nu}}{{\rm d}\tau})|_{\tau=\infty}
 \nonumber\\ 
\mbox{for}\;\; 0\leq\mu\leq1.
\end{eqnarray}
In our computation, close to the lower boundary deep in the atmosphere 
the frequency of low-energy photons can be lower than the local plasma frequency.
In such a case, the O-mode photon has a non-zero imaginary part in the refraction index.
To deal with such a situation, we define an additional opacity $\kappa_{\rm \delta}$ to represent
the opaqueness of the plasma with
$\kappa_{\rm \delta}=\frac{2N_{\rm I}\omega}{\rho c}$, where $N_{\rm I}$ is the imaginary part 
of the refraction index $N$ of the O-mode photons (Rajagopal et al. 1997). 
In the radiative equilibrium assumption,
the total radiative flux is constant, that is,
$\frac{{\rm d}H}{{\rm d}\tau}=0$, 
where $H=\sum_{i=1}^{2}\int{\oint{I_{\nu}^{i}\mu}d\Omega}{\rm d}\nu/(4\pi)$.

\section{Numerical computation}

We use the Feautrier method (Mihalas 1978) to solve 
the radiative transfer equation 
and also adopt the improved Feautrier method 
for reducing numerical round-off errors 
(Rybicki \& Hummer 1991). 
The mean intensity-like and the flux-like quantities are 
defined as
\begin{eqnarray}
P_{\nu}^{i}(\hat{k})=\frac{I_{\nu}^{i}(\hat{k})+I_{\nu}^{i}(-\hat{k})}{2}\,\,\,,\nonumber\\
R_{\nu}^{i}(\hat{k})=\frac{I_{\nu}^{i}(\hat{k})-I_{\nu}^{i}(-\hat{k})}{2}\,\,\,,
\end{eqnarray}
where $\hat{k}$ is the unit wave vector of the out-going photons. 
With the above formulation, the radiative transfer equation becomes 
a second-order differential equation with two boundary conditions.
We choose the gray-atmosphere temperature profile,
 $T^4=\frac{3}{4}T_{\rm e}^4(\tau+\frac{2}{3})$, 
as the first trial profile to start the iteration to obtain the 
radiative transfer equation solution in radiative
equilibrium for a given effective temperature. 
In our computation we choose 20 logarithmically equally-spaced grid points 
in a decade 
of the Thomson optical depth 
for the range from $\tau=10^{-5}$ to $\tau=10^{3}$ and similarly 10 grid points
 per decade in the frequency space for  the photon frequency from
$10^{16}$ to $10^{19}$ Hz. 
Typically specific intensities in about 90 different 
directions in a hemisphere are involved in
 the computation for an arbitrary magnetic field orientation. 
The number of directions is
significantly reduced when the field is perpendicular to 
the stellar surface because of the
axial symmetry. 
The temperature correction is performed with the Uns\"old-Lucy procedure
 (Mihalas 1978).
In a typical computation, about 100 iterations are needed 
to achieve an accuracy at the level of 1\%.

\begin{figure}
\centerline{\psfig{file=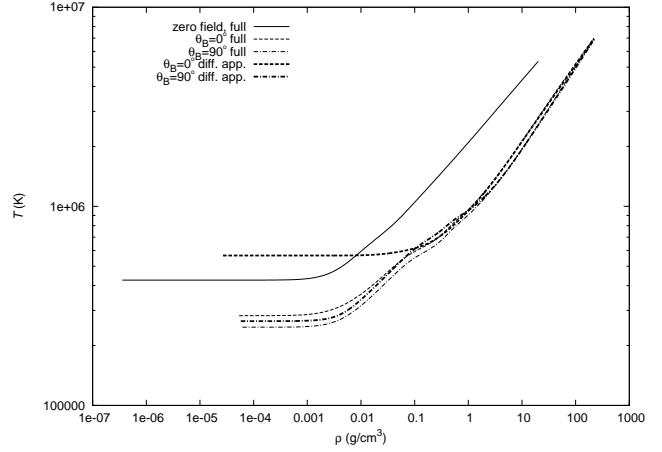,width=6.0cm,angle=-90} }
\caption{Temperature profiles of model atmospheres with
$g^*=10^{14}$ cm/s$^2$, $T_{\rm e}=10^6$ K, and $B=10^{12}$ G (the solid line is for the
case without magnetic fields).
$\theta_{\rm B}$ is the angle between the magnetic field and the surface normal.
}
\label{fig1}
\end{figure}

\begin{figure}
\centerline{\psfig{file=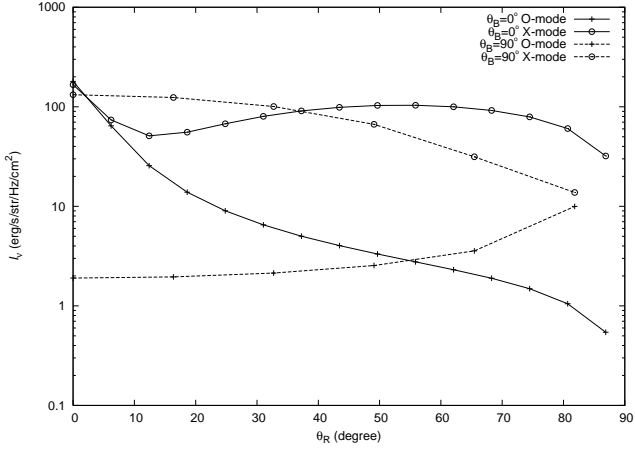,width=6.0cm,angle=-90} }
\caption{The beaming pattern as a function of the photon propagation polar angle
$\theta_{\rm R}$. 
This figure shows the beaming pattern of a $10^{17}$-Hz photon for 
different field orientation and polarization modes.
Parameters used are the same as those in Fig.\ 1
and $\phi_{\rm R}=0^{\circ}$.
}
\label{fig2}
\end{figure}

\begin{figure}
\centerline{\psfig{file=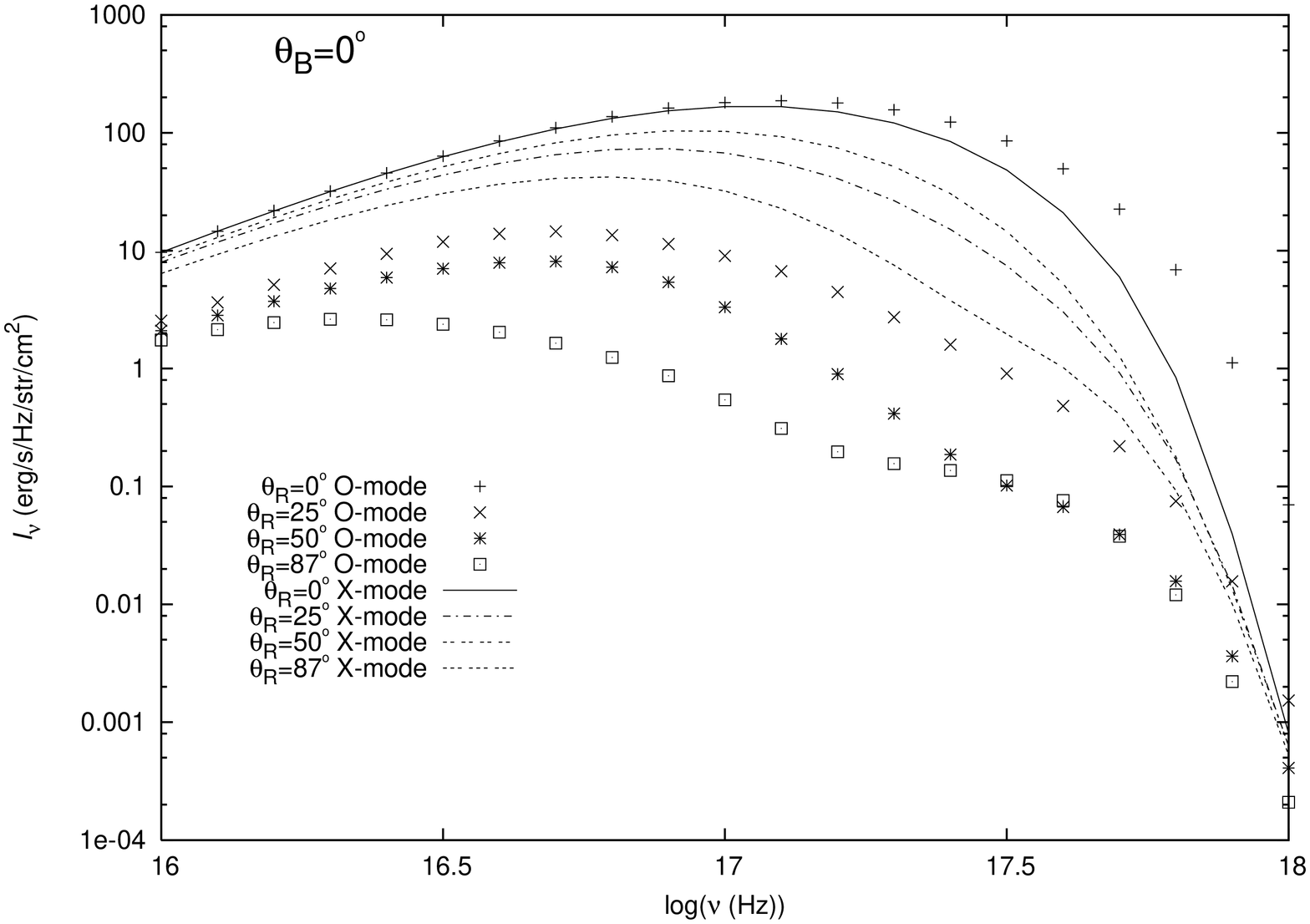,width=6.0cm,angle=-90} }
\centerline{\psfig{file=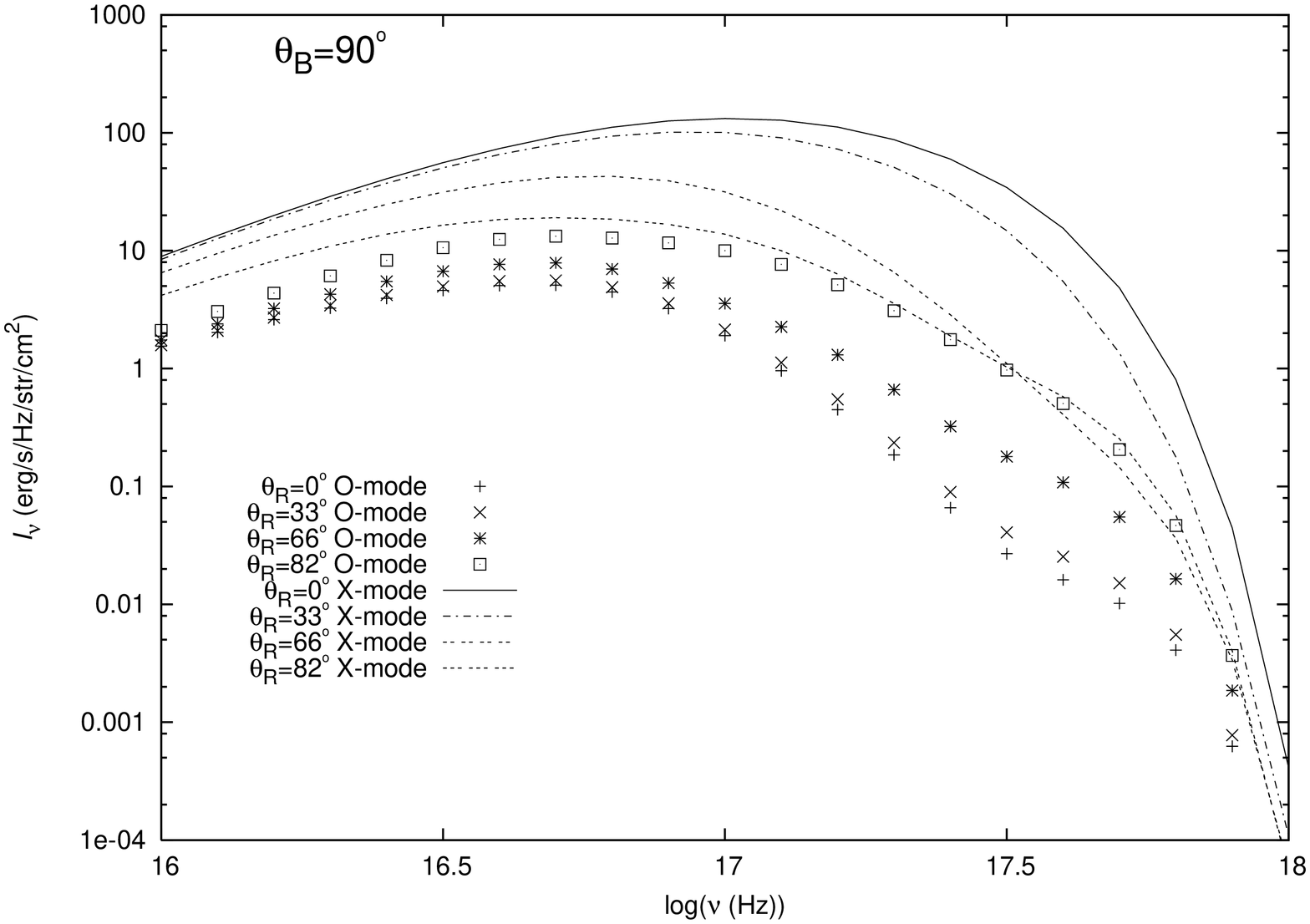,width=6.0cm,angle=-90} }
\caption{Surface emission intensity spectra obtained with the full computation for different 
$\theta_{\rm R}$ and different polarization modes. The upper panel is for the case of
$\theta_{\rm B}=0^\circ$ and the lower for $\theta_{\rm B}=90^\circ$.
Other parameters used are the same as those in Fig.\ 1.
}
\label{fig3}
\end{figure}

\begin{figure}
\centerline{\psfig{file=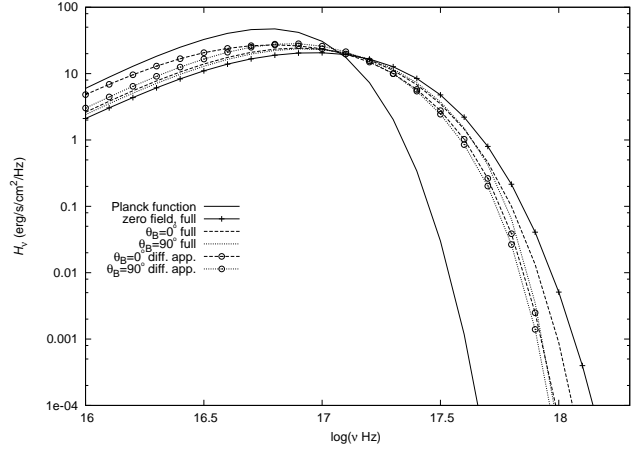,width=6.0cm,angle=-90} }
\caption{Surface emission flux spectra obtained with the full computation
and with the diffusion approximation.
Parameters used are the same as those in Fig.\ 1.
The Planck function of the same effective temperature is shown for comparison.
}
\label{fig4}
\end{figure}

\subsection*{The diffusion approximation}

We also conduct computations with the diffusion approximation (e.g. Shibanov et al. 1992; Ho \& Lai 2001),
in which the specific intensity is approximated as
$I_{\nu}^{i}=\frac{c}{4{\pi}}(u_{\nu}^{i}+\frac{3}{c}\mu F_{\nu}^{i})$ 
at all depth, where $u_{\nu}^{i}$ and $F_{\nu}^{i}$ 
are the energy density and flux density for the $i$-mode respectively.
With this relation the radiative transfer equation for the specific intensity can be turned into that for 
the energy density, which can be solved with much less computing resource. However, it is obvious that in the 
presence of magnetic fields, 
the specific intensity cannot be well approximated by this form (see Fig.\ 2).
The same as in the case of the full computation,
we use the Feautrier method  to solve the radiative transfer equation and
 the Uns\"old Lucy procedure
for the temperature correction iteration.

\section{Results and disscusion}

For a given effective temperature, the computed temperature profiles of the neutron star atmosphere 
with and without the diffusion approximation are plotted in Fig.\ 1. A zero-field temperature profile is also shown for 
comparison. 
Since opacities are generally reduced in the presence of magnetic fields, a magnetized atmosphere is more transparent down
to a denser layer. 
The temperature profiles of magnetized atmospheres show a plateau near $\rho\sim0.1$ g/cm$^3$. 
These plateaus are the manifestation of the different depths at which the O-mode and X-mode photons decouple with matter.
In general, opacities are larger for the O-mode photons and therefore they decouple with matter in a region outer in the atmosphere than
the X-mode photons.
The temperature profile does not differ much for different magnetic field orientations if a full computation without the 
diffusion approximation is performed. For those with the diffusion approximation, the deviation from the full computation
is larger for the case with a smaller $\theta_{\rm B}$.

The beaming pattern, that is, the intensity in different directions of propagation, is complicated.
It is a combination of the effects of limb darkening and the anisotropy due to the presence
of strong magnetic fields. It also depends on the photon frequency and polarization modes
In Fig.\ 2 we show the beaming pattern of the two polarization modes for a $10^{17}$-Hz photon.
Two cases, $\theta_{\rm B}=0^\circ$ and $90^\circ$, are shown. To show the frequency dependence of these patterns,
the intensity spectra in certain propagation directions are shown in Fig.\ 3.
In most cases, the X-mode intensity is generally stronger than the O-mode. 
It is also obviously seen in Fig.\ 2 that the diffusion approximation ansatz,
$I_\nu^i=\frac{c}{4{\pi}}(u_\nu^i+\frac{3}{c}\mu F_\nu^i)$,
is not a good description for radiations in a magnetized atmosphere.

In Fig.\ 4 computed flux spectra are shown. At a given effective temperature, due to the atmospheric effect, that is,
higher energy photons have a smaller opacity, all spectra
are harder than the Planck function of the same effective temperature. 
In the presence of magnetic fields, the spectra get somewhat softer, but still harder than the Planck function.
We particularly note that the difference between the spectra computed with and without the diffusion approximation
 is noticeable and is distinguishable with the current observation accuracy.   
Those with the diffusion approximation are generally softer than those without. The effective temperature
will be over-estimated when fitting the observed spectrum with the model computed with the diffussion approximation. 
We therefore urge readers to be cautious when applying the diffusion approximation.
 
The results discussed in this paper are those for a local patch at the neutron
star surface. To compare with observations, even for phase-resolved spectra, a
global spectrum is needed, which is a combination of emission from different 
patches all over the neutron star surface. Gravitational light bending and 
redshift should be also considered. Since contributions come from different
part of the stellar surface, different magnetic field orientations 
relative to the local surface normal are inevitable. 
Although such a computation has yet to be performed and knowledge 
of the temperature distribution over the whole stellar surface is required,
our results (Fig.\ 2 and Fig.\ 3) already indicate
that the global spectrum will be polarized to a significant degree in the
extraordinary mode. More quantitative and detailed results will be obtained
with a more comprehensive computation in the future. 

\vskip 0.4cm

\begin{acknowledgements}
This work was supported by the National Science Council of the Republic of China with the
grant NSC 94-2112-M-007-002.
\end{acknowledgements}

                  \clearpage

\end{document}